\documentclass[11pt,english]{article}
\usepackage[T1]{fontenc}
\usepackage[utf8]{luainputenc}
\usepackage{geometry}
\usepackage{amstext}
\usepackage{amssymb}
\usepackage{setspace}
\usepackage{esint}
\makeatletter
\usepackage{babel}

\makeatother

\usepackage{babel}
\begin{document}
\title{Proposed evolution in Marolf-Maxfield toy model obtained through correspondence
to spontaneous collapse theory }
\author{{\small{}Merav Hadad{} }\thanks{{\small{}{}meravha@openu.ac.il }}}
\maketitle
\begin{abstract}
We consider a general correspondence between wave function collapse
and evolution through wormholes, which presumably occurs whenever
a black hole evaporates after Page time. We use this correspondence
in order to explore a possible evolution from the Hartle-Hawking state
to one of the superselection sectors for a topological model. The
model considered is the Marolf-Maxfield (MM) topological toy model
for 2D gravity which gives the full spectrum of boundary theories.
By equating the MM topological toy model to the Bonifacio model of
spontaneous collapse, an equation for the evolution of a matrix element
due to a generator of a parameter in MM model is obtained. 
\end{abstract}

\section{Introduction}

Recently a unitary black hole evaporation was obtained. This was done
by considering a quantum extremal surface (QES) which was created
inside the black hole. It turns out that the QES mysteriously enables
particles deep inside the black hole to no longer be part of the hole,
but rather part of the radiation. One finds that in order to understand
this formalism, one needs to include summing over wormhole geometries'
connecting boundaries and that these wormholes enable the entropy
of Hawking radiation to follow the expected time dependence \cite{netta}.
This evolution is unique and can not be obtained by using ordinary
equations of motion. A topological toy model for 2D gravity, such
as that of Marolf-Maxfield (MM)\cite{with rut}, can consider the
presence of spacetime wormholes which connect boundaries. Although
these models are purely topological and can not describe time evolution,
they give the full spectrum of boundary theories. Assuming wormhole
and other topological fluctuations lead to a unique evolution in gravity,
we suggest to explore this evolution by considering the correspondence
which exists between two processes: \emph{evaporation of black holes}
after Page time on the gravity side and \emph{collapse of the wave
function} on the quantum side.

Although black hole evaporation and wave function collapse occur at
very different scales and are due to very different theories, a similar
mathematical structure between the two scenarios could be expected
not only in light of \cite{Susskind:2016jjb,Susskind:2017ney} which
suggest that general relativity may already be a quantum theory, but
also due to the existence of similarities between the two scenarios. 

To start with they both have a kind of\emph{ apparent loss of information.
}Black holes look as if they lose their information by Hawking radiation
during their evaporation. According to Copenhagen interpretation,
wave functions lose their information when collapsing to one of the
eigenstates, since all information on the original function is lost
during a measurement.\emph{}\footnote{Note that though we consider the Copenhagen interpretation, the approach
	that a quantum state represents only knowledge and “facts” can only
	exist relative to the observer as proposed by Brukner \cite{Brukner}
	is more relevant in our correspondence, since in this approach the
	loss of information is only an apparent property.}

Moreover they both have a kind of \emph{action at a distance}. Black
holes look as if they act at a distance since they must connect Hawking
radiation at infinity to the interior of the black hole in order for
the black hole to be able to evaporate after Page time. Wave functions
look as if they act at a distance whenever a collapse takes place
since we expect a wave function to collapse all at once in all the
points in the universe. 

Finally, the most subtle and surprising resemblance is related to
the suggestion that both scenarios involve a kind of decoherence.
In general decoherence can be viewed as the loss of information from
a system into an environment. When collapsing of a wave function
is considered, decoherence provides a framework for an apparent collapse.
In this case a quantum system begins to obey classical probability
rules due to suppression of interference terms after interacting with
its environment. This leads to a known outcome in quantum theory:
observers can not consider the result of an experiment as a measurement
of previously unknown value. On the other hand when evaporation of
black hole through a wormhole is considered a different decoherence
is obtained. As was first noted by Coleman \cite{Coleman}: \textquotedbl topological
fluctuations can not lead to an observable loss of quantum coherence.\textquotedbl{}
Considering wormholes as topological fluctuation one finds that a
loss of quantum coherence can not be obtained if black hole evaporates
through wormholes and in this case, observers may consider the result
of an experiment to be a measurement of a previously unknown value.
To emphasize, whereas observers cannot consider the result of an experiment
as a measurement of previously unknown value when a collapsing wave
function is considered, they can do so when an evaporating black hole
or any other evolution through wormholes is considered. 

This important difference between the decoherence of a collapsing
wave function and the decoherence due to topological fluctuations
is crucial in our analogies, since it limits the relevant models of
wave function collapse. In other words, in order to obtain mathematical
correspondence between evolution due to topological fluctuations and
wave functions collapse we must consider only those mathematical models
of collapsing wave functions which cause non-observable decoherence. 

The purpose of the present work is to use this apparent correspondence
in order to obtain an equation for the evolution of the density matrix
in MM topological toy model for 2D gravity. Marolf and Maxfield constructed
a Hartle-Hawking state, after computing the full spectrum of the associated
boundary theories and considering the presence of spacetime wormholes
connecting these boundaries. However, there is no evolution in MM
model, since this model is a topological model and there is no time.
As a matter of fact, if one tries to consider evolution through
wormholes, one may expect a significant
change in the equations of motion, such as apparent
loss of information and action at a distance. 

In this short paper we suggest obtaining the differential equation
for the evolution of the density matrix through wormholes, by equating
it to a model of collapse that has similar mathematical and physical
characteristics. We consider models of collapse which uses the Lindblad
equation \cite{review collapse}. However, as expected, the observation
that evolution through wormholes does cause non-observable decoherence
restricts the Lindblad equation as well as the possible expected models
which produce the collapse. This leads us to Bonifacio model \cite{Bonifacio} which does not cause observable decoherence. Moreover, when MM model
is considered one finds that the two models have the same mathematical
structure: the distribution considered in both Bonifacio model and
MM-model is the Poisson distribution. This enables us to obtain an
equation of the \textquotedbl evolution\textquotedbl{} of an element
of the density matrix in MM model.

This paper is organized as follows: Section 2 briefly summarizes recent relevant concepts on black hole evaporation and MM model. Section
3 briefly review relevant spontaneous collapse theories and the Bonifacio model. Section
4 obtains the density matrix for the MM model and its evolution
equation. Then, by equating the two density matrix evolution equations,
the evolution of a matrix element due to a generator of a parameter
in MM model is obtained. Section 4 is a summary and conclusion. 

\section{Brief review of the BH evaporation and Marolf-Maxfield model}

Recent years have seen significant progress towards the resolution
the puzzle of black hole evaporation \cite{netta,Penington,key-14,key-15,key-16,key-17,key-18,Goto}.
This puzzle comes from semi-classical computations that imply that
black hole evaporation is non-unitary, in stark contrast with the
principles of quantum mechanics \cite{Hawking non unitarity}. To
see this, consider forming a black hole from a pure state with enough
energy in a compact region of a quantum gravity system. Such a black
hole seems to evolve from a pure state to a mixed thermal state through
Hawking radiation. This amounts to a loss of information and is incompatible
with unitary time evolution. Moreover, it turns out that if black
hole evaporation is a unitary process, the entanglement entropy between
the outgoing radiation and the quantum state associated to the remaining
black hole is expected to follow the Page curve \cite{Page 1,Page 2}.
According to the Page curve initially the black hole entropy grows,
but at some point in time - the Page time - the entropy has to stop
rising and start dropping. The reversal at the Page time would have
to occur roughly halfway through the process, and not at the very
end of the evaporation. Thus, although it was natural to expect this,
it turns out that quantum gravity effects could not restore unitarity
if they occur only towards the very end of the evaporation.

Recently an explicit description of evaporating black holes was obtained
\cite{netta} and succeeded in obtaining the expected Page curve.
The Page curve was obtained for a double sided black hole in two dimensional
Jackiw-Teitelboim (JT) gravity with conformal matter, which is allowed
to evaporate into a non-gravitational reservoir coupled to one side
of the black hole \cite{netta,Penington,Chen}. This calculation applied
the Engelhardt-Wall prescription for computing holographic entanglement
entropy \cite{Wall =000026 neta}. According to this prescription
one can compute the entanglement entropy of a boundary subregion B
by considering all possible co-dimension two surfaces $\sigma_{B}$
which are homologous to the boundary subregion $B$. For each $\sigma_{B}$
one can define the generalized entropy 
\begin{equation}
S_{gen}(\sigma_{B})=\frac{A(\sigma_{B})}{4G_{N}}+S_{bulk}(\Sigma_{B})\label{eq: entropy-1}
\end{equation}
where $\Sigma_{B}$ is a bulk codimension-one surface bounded by $\sigma_{B}$
and $B$, i.e.,$\partial\Sigma_{B}=\sigma_{B}\bigcup B$. As we can
see, the entanglement entropy of a boundary subregion $B$ involves
two terms: the Bekenstein-Hawking area term in the Hubeny-Rangamani-Ryu-Takayanagi
prescription \cite{RT,RT-H} and the quantum corrections in the form
of the von Neumann entropy of the quantum fields on $\Sigma_{B}$.
According to the prescription, in order to compute the entanglement
entropy of a boundary subregion B, one then extremizes the generalized
entropy over all surfaces $\sigma_{B}$ that are homologous to B.
Surfaces $\sigma_{B}$ that extremize the generalized entropy are
referred to as quantum extremal surfaces (QES). It turn out that during
the evaporation, a black hole creates a QES. After the Page time the
QES is located just inside the horizon of the black hole and mysteriously
enables particles deep inside the black hole to no longer be part
of the hole, but rather part of the radiation. This inner core of
radiation is an “island”. 

One of the ways to obtain the formalism that led to the
Page curve is summing over wormholes. Wormholes can be viewed as an outcome of the 
Hilbert space interpretation that includes states of closed ‘baby’ universes
that propagate between distinct asymptotic boundaries. The baby universe
Hilbert space of any theory can be constructed by acting on a state
with no-boundaries, typically called the Hartle-Hawking state, with
boundary creation operators. Alpha-states are the special states in
this space that are eigenfunctions of all such operators, leading
to inner products in the baby universe Hilbert space.

To summarize,
these recent works suggest that a unitary black hole evaporation can
be obtained and that this evaporation uses wormholes geometries with
connecting boundaries.

\subsection*{Marolf-Maxfield model for Hartle-Hawking no-boundary state }

Marolf and Maxfield \cite{with rut} considered a topological model
with asymptotic boundaries and computed the full spectrum of the associated
boundary theories by considering the presence of spacetime wormholes
connecting these boundaries. They obtained a Hartle-Hawking state
with Poisson distribution for the number of connected components. 

Here we point out only the relevant features of their model. 

The Marolf-Maxfield \cite{with rut} topological model is a simple
2-dimensional model of the bulk with asymptotic AdS Hilbert space
that is represented by a random variable $Z$ with nonnegative integer
values. By using the Taylor series of the generating function $e^{uZ}$
and identifying the parameter $\lambda$ as the sum over connected
compact surfaces, they obtain $Z$ as a Poisson random variable with
mean $\lambda$. They characterize the state $\left|Z=d\right\rangle $
as a spacetime with $d$ connected components, and define an annihilation
operator acting to shift functions of $Z$: $a\left|f(Z)\right\rangle =\sqrt{\lambda}\left|f(Z+1)\right\rangle $,
so that $\hat{Z}=N=a^{\dagger}a,\text{and }$$a\left|Z=0\right\rangle =0$.
This enables Marolf and Maxfield to obtain the state $\left|Z=d\right\rangle $
as:
\begin{equation}
\left|Z=d\right\rangle =\frac{1}{\sqrt{d!}}\left(a^{\dagger}\right)^{d}\left|Z=0\right\rangle \label{eq:Z=00003DD}
\end{equation}
and the Hartle-Hawking no-boundary state as:
\begin{equation}
\left|HH\right\rangle =e^{\sqrt{\lambda}a^{\dagger}}\left|Z=0\right\rangle .\label{eq:Hartel_Hawking}
\end{equation}
This causes the number operator to follow a Poisson distribution in
a coherent state. Marolf and Maxfield noted that the appearance of
the Poisson distribution can be understood from the result that all
connected components of spacetime contribute the same amplitude after
summing over genus, independent of the number of boundaries.

\section{Brief review of the spontaneous collapse theory and Bonifacio model}

Models of spontaneous wave function collapse \cite{QM collapse 1,QM collapse 2}
were formulated as a response to the measurement problem in quantum
mechanics \cite{QM collapse 3}. The fundamental idea is that the unitary
evolution of the wave function describing the state of a quantum system
is approximate. In collapse theories, the Schrödinger equation is
supplemented with additional nonlinear and stochastic terms (spontaneous
collapses) which localize the wave function in space. These additional
stochastic terms change the evolution equation of the density matrix
and give the Lindblad equation \cite{review collapse}:

\begin{equation}
\frac{\partial}{\partial t}\rho(t)=-\frac{i}{\hbar}\left[H,\rho(t)\right]-\frac{\lambda}{2}\left[A,\left[A,\rho(t)\right]\right]\label{in time-master equation-1-1-2}
\end{equation}
The second term on the RHS stochastically drives the state vector
toward one of the eigenstates of the operator $\hat{A}$ with a probability
equal to the Born rule \cite{QM collapse 4,QM collapse 5}. 

However, it is interesting to note that this does not always lead to an observable decoherence \cite{Banks and...,Giddings =000026 Strominger}. This interesting and
surprising property of a collapse theory is very important for our correspondence, since it is
in agreement with the observation \cite{Coleman} that quantum decoherence
due to information loss to baby universes is not experimentally observable.\footnote{I want to thank Don Marolf for clarifying this important point for
me.}

\subsection*{Bonifacio model forspontaneous collapse theory}
The Bonifacio model presents an interesting way to derive the structure
of a collapse that does not lead to an observable decoherence \cite{Bonifacio}.
This is done as follows. According to this model, instead of the continuous
unitary evolution of the density matrix; i.e.:

\begin{equation}
\rho(t+\tau)=exp(-\frac{i}{\hbar}H\tau)\rho(t)exp(\frac{i}{\hbar}H\tau)\label{in time-1-1}
\end{equation}
one assumes that a change of the system occurs with a certain probability:
specifically, that there is a probability $p_{n}(\tau_{0},\tau)$
such that $n$ random time shifts by $\tau_{0}$ lead to a total time
shift of $\tau$. The density matrix then satisfies a probabilistic
evolution equation:

\begin{equation}
\rho(t+\tau)=\sum_{n=0}^{\infty}p_{n}(\tau_{0},\tau)exp(-\frac{i}{\hbar}Hn\tau_{0})\rho(t)exp(\frac{i}{\hbar}Hn\tau_{0}).\label{in time-probability-1}
\end{equation}
 In the limit $\tau_{0}\rightarrow0$ this yields the continuous equation
(\ref{in time-1-1}). Next assuming a Poisson distribution

\begin{equation}
p_{n}(\tau_{0},\tau)=\frac{1}{n!}\left(\frac{\tau}{\tau_{0}}\right)^{n}e^{-\tau/\tau_{0}},\label{Poisson-1}
\end{equation}
one obtains an interesting result: the master equation takes the form

\begin{equation}
\frac{\partial}{\partial t}\rho(t)=\frac{1}{\tau_{0}}\left(exp(-\frac{i}{\hbar}H\tau_{0})\rho(t)exp(\frac{i}{\hbar}H\tau_{0})-\rho(t)\right).\label{in time-master equation-2}
\end{equation}
We consider the parameter $\tau_{0}$ to be small, since this indicates
the time scale where deviations from continuous evolution become apparent.
Expanding in $\tau_{0}$ the master equation takes the form:

\begin{equation}
\frac{\partial}{\partial t}\rho(t)=-\frac{i}{\hbar}\left[H,\rho(t)\right]-\frac{\tau_{0}}{2\hbar^{2}}\left[H,\left[H,\rho(t)\right]\right].\label{in time-master equation-1-2}
\end{equation}
As pointed out by Anastopoulos and Hu \cite{Anastopoulos and Hu},
these considerations are in fact more general, and the semi-group
equation \cite{Mavromatos  and Sarkar} follows quite generally for
any physical quantity and its corresponding generator. For example,
the above results remain valid if time is replaced by position and
the time translation generator $H$ by the momentum $p$.

Since we construct a correspondence between Bonifacio
collapse model and black hole evaporation, we will need to write the
Bonifacio model in a unitless form. For a unitless $\lambda$ and
its unitless generator\footnote{For example $B\equiv H\tau_{0}/\hbar$ , $\lambda=\tau/\tau_{0}$
, and $\lambda_{0}=t/\tau_{0},$} $B$, the standard evolution for a density matrix $\rho(\lambda_{0}+\lambda)=exp(-iB\lambda)\rho(\lambda_{0})exp(iB\lambda)$
takes the form: 

\begin{equation}
\rho(\lambda_{0}+\lambda)=\sum_{\text{d}=0}^{\infty}p_{d}(\lambda)exp(-iBd)\rho(\lambda_{0})exp(iBd)\text{.}\label{BU-probability-1-2}
\end{equation}
 For the choice of a Poisson distribution
\begin{equation}
p_{d}(\lambda)=\frac{1}{d!}\left(\lambda\right)^{d}e^{-\lambda}\label{poisson for lamda}
\end{equation}
the master equation takes the form

\begin{equation}
\frac{\partial}{\partial\lambda}\rho(\lambda)=exp(-iB)\rho(\lambda)exp(iB)-\rho(\lambda).\label{no unit-master equation-2-1-2-2}
\end{equation}
After expansion in the presumably small parameter $B$, the evolution
becomes apparent:$\frac{\partial}{\partial\lambda}\rho(\lambda)\simeq-i\left[B,\rho(\lambda)\right]-\frac{1}{2}\left[B,\left[B,\rho(\lambda)\right]\right]$
and again the new term drives the state vector toward one of the eigenstates
of the operator $\hat{B}$. 

\section{Constructing the correspondence}

In this section we equate Bonifacio model to Marolf-Maxfield's topological
model. The first subsection gives the Hartle-Hawking no-boundary state
density matrix for MM model and derives its \textquotedbl evolution\textquotedbl{}
equation. The second subsection equates the two density matrix equations:
the master equation and Hartle-Hawking evolution equation. This gives
the condition for them to be the same. 

\subsubsection*{\textquotedbl Evaporating\textquotedbl{} Hartle-Hawking state}

In order to construct a correspondence between the MM model and the Bonifacio model, it is necessary to identify the
rate of change of the Hartle-Hawking density matrix before equating
it to the unitless master equation (\ref{no unit-master equation-2-1-2-2}). 

Using eq. (\ref{eq:Hartel_Hawking}), one finds that the Hartle-Hawking
density matrix for the MM model can be written as 
\begin{equation}
\hat{\rho}=\xi^{-1}\left|HH\right\rangle \left\langle HH\right|=e^{-\lambda}\sum_{m=0}^{\infty}\sum_{n=0}^{\infty}\frac{\lambda^{m/2}}{\sqrt{m!}}\frac{\lambda^{n/2}}{\sqrt{n!}}\left|Z=n\right\rangle \left\langle Z=m\right|.\label{eq:density metric}
\end{equation}
This density matrix has a Poisson distribution, as does the Bonifacio
model for spontaneous collapse theory. Note that although in the original Bonifacio model the parameter $\lambda$
is a unitless time coordinate $\lambda=\tau/\tau_{0}$, according
to \cite{QM collapse 4} this can be generalized to any physical quantity
and its corresponding generator. 

In order to construct an equation similar to (\ref{no unit-master equation-2-1-2-2}),
the rate of change of the Hartle-Hawking density matrix should also
be calculated with respect to the parameter $\lambda$. Using eq.
(\ref{eq:density metric}) one finds (see Appendix) that: 
\begin{equation}
\frac{d}{d\lambda}\hat{\rho}=-\hat{\rho}+\frac{1}{2}\lambda^{-1/2}\left(a^{\dagger}\hat{\rho}+\hat{\rho}a\right).\label{eq:eq density metric GR}
\end{equation}

This gives the equation for the rate of \textquotedbl evolution\textquotedbl{}
of the Hartle-Hawking density matrix.

\subsubsection*{Equating MM Hartle-Hawking evolution density matrix to the master equation. }

Next we require that the MM Hartle-Hawking evolution density matrix agree with
the master equation obtained in the Bonifacio model. In other words,
we equate the terms in the unit-less\textbf{ }master equation for
the Bonifacio model (\ref{no unit-master equation-2-1-2-2}) and the
rate of change of Hartle-Hawking density matrix for the MM model (\ref{eq:eq density metric GR}). This gives a connection between the two generating operators $B$
and $a$: 

\begin{equation}
exp(-iB)\rho(\lambda)exp(iB)=\frac{1}{2}\lambda^{-1/2}\left(a^{\dagger}\rho(\lambda)+\rho(\lambda)a\right).\label{connecting a and B-1}
\end{equation}
This is a strange connection. Whereas the operator $a$ promotes the
number of the disconnected components of spacetime, the operator $B$
promotes $\lambda$, which is a free parameter in the theory. In order
to try and understand this connection we use eq. (\ref{eq:Z=00003DD})
for spacetimes with $d$ connected components, and obtain that in
order that the black hole density matrix stochastically drive the
state vector toward one of the eigenstates, the operator $B$ \textquotedbl evolves\textquotedbl{}
the density matrix element $\left|Z=n\right\rangle \left\langle Z=m\right|$
to:
\[
exp(-iB)\left|Z=n\right\rangle \left\langle Z=m\right|exp(iB)=
\]
\begin{equation}
=\frac{1}{2}\left(\left|Z=n+1\right\rangle \left\langle Z=m\right|+\left|Z=n\right\rangle \left\langle Z=m+1\right|\right).
\end{equation}
Indeed, the operator $B$ changes the density matrix element $\left|Z=n\right\rangle \left\langle Z=m\right|$
to a \textquotedbl superposition of density matrix elements\textquotedbl{}
with a different number of disconnected spacetimes.

\section{Summary and discussion}

This note obtains an \textquotedbl evolution\textquotedbl{} of the
density metrics in MM model by considering a possible correspondence
between collapsing wave functions and evolution through wormholes
(as occur when black holes evaporate after Page time). By equating
the Marolf-Maxfield topological toy model to the Bonifacio model for
spontaneous collapse theory, a condition for their evolution equation
to be the same is obtained. This gives the evolution that should be
expected from the matrix element $\left|Z=n\right\rangle \left\langle Z=m\right|$ in
order for the Hartle-Hawking state to evolve from Hartle-Hawking state to one of the superselection
sectors. This \textquotedbl evolution\textquotedbl{} is describe
by $exp(-iB)$, where $B$ is a generator of the parameter $\lambda$
in MM topological toy model\footnote{So that the operator $exp(-\frac{i}{\hbar}B)$ shifts the value of
$\lambda$ to $\lambda+1$} 

By construction, we obtained that the evolution in MM-model should
be related to changes in the parameter in the theory. This may look
surprising at first glance, but looks reasonable if we consider that
MM-model have the following property: different values of an apparently
free parameter turn out to describe different states of the same theory.
In this case, changes in the free parameter can be consider an evolution
from one state to the another, as one may expects when evolution take
place. 

This correspondence could shed some light on the nature of collapse. For example it may clarify the origin of the Bonifacio model. In order to see that, note that in the MM
model
the Hartle-Hawking state involves spacetimes with $d$ connected components
$\left|Z=d\right\rangle $. If our correspondence is accurate,
the Bonifacio model corresponds to the
distribution of \textquotedbl$d${} connected components\textquotedbl{}
whose nature is not clear, but which somehow enables transmission
of information without breaking causality in the same way that wormholes
do. This would present an exciting avenue for further research on the nature of collapse
which is still not properly understood.

\textbf{Acknowledgments:} This research was supported by The Open
University of Israel's Research Fund (grant no. 510503).

\section*{Appendix: Varying Hartle-Hawking density matrix with respect to $\lambda$}

Since

\begin{equation}
\hat{\rho}=\xi^{-1}\left|HH\right\rangle \left\langle HH\right|=
\end{equation}

\begin{equation}
=\xi^{-1}e^{\sqrt{\lambda}a^{\dagger}}\left|Z=0\right\rangle \left\langle Z=0\right|e^{\sqrt{\lambda}a}=
\end{equation}

\begin{equation}
=e^{-\lambda}\sum_{m=0}^{\infty}\sum_{n=0}^{\infty}\frac{\lambda^{m/2}}{m!}\frac{\lambda^{n/2}}{n!}\left(a^{\dagger}\right)^{n}\left|Z=0\right\rangle \left\langle Z=0\right|a^{m}\label{eq:ro for derivativ}
\end{equation}
\begin{equation}
=e^{-\lambda}\sum_{m=0}^{\infty}\sum_{n=0}^{\infty}\frac{\lambda^{m/2}}{\sqrt{m!}}\frac{\lambda^{n/2}}{\sqrt{n!}}\left|Z=n\right\rangle \left\langle Z=m\right|,
\end{equation}
the derivation of eq. (\ref{eq:ro for derivativ}) with respect to
$\lambda$ is

\[
\frac{d}{d\lambda}\hat{\rho}=-e^{-\lambda}\sum_{m=0}^{\infty}\sum_{n=0}^{\infty}\frac{\lambda^{m/2}}{m!}\frac{\lambda^{n/2}}{n!}\left(a^{\dagger}\right)^{n}\left|Z=0\right\rangle \left\langle Z=0\right|a^{m}+
\]

\[
+\frac{e^{-\lambda}}{2}\sum_{m=0}^{\infty}\sum_{n=0}^{\infty}m\frac{\lambda^{m/2-1}}{m!}\frac{\lambda^{n/2}}{n!}\left(a^{\dagger}\right)^{n}\left|Z=0\right\rangle \left\langle Z=0\right|a^{m}+
\]
\[
+\frac{e^{-\lambda}}{2}\sum_{m=0}^{\infty}\sum_{n=0}^{\infty}n\frac{\lambda^{m/2}}{m!}\frac{\lambda^{n/2-1}}{n!}\left(a^{\dagger}\right)^{n}\left|Z=0\right\rangle \left\langle Z=0\right|a^{m}
\]

\[
\frac{d}{d\lambda}\hat{\rho}=-\hat{\rho}+
\]

\[
+\frac{e^{-\lambda}}{2}\sum_{n=0}^{\infty}n\frac{\lambda^{n/2-1}}{n!}\left(a^{\dagger}\right)^{n}\left|Z=0\right\rangle \left\langle HH\right|+
\]
\[
+\frac{e^{-\lambda}}{2}\sum_{m=0}^{\infty}m\frac{\lambda^{m/2-1}}{m!}\left|HH\right\rangle \left\langle Z=0\right|a^{m}.
\]
Since $m=0$ and $n=0$ does not contribute, we can state the summation
from $m=1$ and $n=1$ 

\[
\frac{d}{d\lambda}\hat{\rho}=-\hat{\rho}+
\]
\[
+\lambda^{-1/2}\frac{e^{-\lambda}}{2}a^{\dagger}\sum_{n=1}^{\infty}\frac{\lambda^{(n-1)/2}}{(n-1)!}\left(a^{\dagger}\right)^{n-1}\left|Z=0\right\rangle \left\langle HH\right|+
\]
\[
+\lambda^{-1/2}\frac{e^{-\lambda}}{2}\sum_{m=1}^{\infty}\frac{\lambda^{(m-1)/2}}{(m-1)!}\left|HH\right\rangle \left\langle Z=0\right|a^{m-1}a.
\]
We rename $n-1\rightarrow n$ and $m-1\rightarrow m$ thus the summation
start again from $0$:

\[
\frac{d}{d\lambda}\hat{\rho}=-\hat{\rho}+
\]
\[
+\lambda^{-1/2}\frac{e^{-\lambda}}{2}a^{\dagger}\sum_{n=0}^{\infty}\frac{\lambda^{n/2}}{(n)!}\left(a^{\dagger}\right)^{n}\left|Z=0\right\rangle \left\langle HH\right|+
\]
\[
+\lambda^{-1/2}\frac{e^{-\lambda}}{2}\sum_{m=0}^{\infty}\frac{\lambda^{(m)/2}}{(m)!}\left|HH\right\rangle \left\langle Z=0\right|a^{m}a
\]

\[
\frac{d}{d\lambda}\hat{\rho}=-\hat{\rho}+\frac{1}{2}\lambda^{-1/2}\left(a^{\dagger}\hat{\rho}+\hat{\rho}a\right).
\]

\end{document}